\documentclass[twocolumn,aps,prb,groupedaddress,superscriptaddress,amsmath,floatfix,amssymb,showpacs,noeprint,english]{revtex4-2}
\usepackage{graphicx}
\usepackage{dcolumn}
\usepackage{bm}
\usepackage[T1]{fontenc}
\usepackage{mathrsfs}
\usepackage{amsbsy}
\usepackage{amstext}
\usepackage{amssymb}
\usepackage{setspace}
\usepackage{esint}
\usepackage{xcolor,colortbl}
\usepackage{float}
\usepackage[colorlinks,citecolor=blue,urlcolor=blue,linkcolor=blue,hypertexnames=true]{hyperref} 
\usepackage{babel}
\usepackage{booktabs}
\usepackage{tikz}

\begin{document}

\title{Predicting the onset of quantum synchronization using machine learning}

\author{F. Mahlow}
\email{f.mahlow@unesp.br}
\affiliation{Faculty of Sciences, UNESP - S{\~a}o Paulo State University, 17033-360 Bauru-SP, Brazil}
	
\author{B. Çakmak}
\affiliation{Department of Physics, Farmingdale State College—SUNY, Farmingdale, NY 11735, USA}
\affiliation{College of Engineering and Natural Sciences, Bahçeşehir University, Beşiktaş, Istanbul 34353, Turkiye}

\author{G. Karpat}
\affiliation{Department of Physics, Faculty of Arts and Sciences, İzmir University of Economics,  İzmir, 35330, Turkey}

\author{\.{I}. Yal\c{c}\i nkaya}
\affiliation{Department of Physics, Faculty of Nuclear Sciences and Physical Engineering, Czech Technical University in Prague, B\v{r}ehov\'a 7, 115 19 Praha 1-Star\'e M\v{e}sto, Czech Republic}

\author{F. F. Fanchini}
\affiliation{Faculty of Sciences, UNESP - S{\~a}o Paulo State University, 17033-360 Bauru-SP, Brazil}
\affiliation{QuaTI - Quantum Technology \& Information, 13560-161 São Carlos-SP, Brazil}

\date{\today}

\begin{abstract}

We have applied a machine learning algorithm to predict the emergence of environment-induced spontaneous synchronization between two qubits in an open system setting. In particular, we have considered three different models, encompassing global and local dissipation regimes, to describe the open system dynamics of the qubits. We have utilized the $k$-nearest neighbors algorithm to estimate the long-time synchronization behavior of the qubits only using the early time expectation values of qubit observables in these three distinct models. Our findings clearly demonstrate the possibility of determining the occurrence of different synchronization phenomena with high precision even at the early stages of the dynamics using a machine learning-based approach. Moreover, we show the robustness of our approach against potential measurement errors in experiments by considering random errors in the qubit expectation values, initialization errors, as well as deviations in the  environment temperature. We believe that the presented results can prove to be useful in experimental studies on the determination of quantum synchronization.

\end{abstract}

\maketitle

\section{Introduction}

Machine learning is a rapidly growing field of research that involves the use of computational algorithms to estimate complex functions using large amounts of available data. These functions can then be used to make some predictions and identify patterns in given datasets. The main difference between the machine learning approach and other statistical models is that it allows a computer program to improve its performance or learn, without the need for explicit programming \cite{Schuld2018}. In recent years, machine learning methods have achieved remarkable success in a wide range of applications, including natural language processing, image recognition, drug discovery, and finance. Machine learning algorithms have been applied in many areas, including computer science, medicine, biology, and even social sciences \cite{Jordan2015}. In physics, machine learning techniques have been extensively used in various research fields, encompassing cosmology, particle physics, condensed matter physics, and quantum computing \cite{Carleo2019}.

Synchronization is a widespread phenomenon that occurs across many different systems, from natural systems like the beating of a heart or flashing of fireflies to social systems such as the behavior of a crowd. Physical systems can exhibit synchronous behavior in two ways, namely, forced and spontaneous. When two or more systems are forced to oscillate in unison by an outside influence, such as the regulation of heart rate by an external pacemaker through electrical pulses, this is known as forced synchronization. On the contrary, spontaneous synchronization takes place when two or more systems naturally synchronize in the absence of any external agent. Synchronization in classical systems has been studied in a variety of contexts over the past few decades with many interesting outcomes~\cite{Pikovsky2001, Arenas2008,Osipov2007}. Consequently, the study of this universal phenomenon has been extended to the quantum domain.

Nevertheless, it needs to be emphasized that synchronization is a term of wide scope that can indeed be interpreted in multiple ways~\cite{Mari2013, Li2017, Galve2017, Roulet2018a, Giorgi2019}. Forced synchronization due to an external drive, also known as entrainment, has been studied in driven dissipative oscillators~\cite{Zhirov2008, Zhirov2009}, van der Poll oscillators~\cite{Lee2013,Walter2014,Sonar2018} and spin-boson type models~\cite{Goychuk2006}. On the other hand, environment-induced spontaneous synchronization has also been broadly explored, for example, in harmonic oscillators~\cite{Giorgi2012,Manzano2013,Manzano2013a,Benedetti2016}, van der Poll oscillators~\cite{Lee2014,Walter2015}, optomechanical arrays~\cite{Heinrich2011,Ludwig2013}, cold ions in microtraps~\cite{Hush2015}, atomic lattices~\cite{Cabot2019}, spins coupled to common~\cite{Orth2010,Giorgi2013,Bellomo2017} and local reservoirs~\cite{Giorgi2016}, collectively dissipating few-body atom systems~\cite{Karpat2020}, collision models~\cite{Karpat2019,Karpat2021,Li2023a} and more recently, in many others~\cite{Schmolke2022,Xiao2023,Krithika2022,Eshagi-sani2020,Qiao2020,Li2023,Li2022,Wachtler2023,Cattaneo2021,Cabot2021,Buca2022,Impens2023,Sterba2023,Kalit2021}.

In the last few years, machine learning techniques have started to be employed in physical systems, where synchronous behavior between their constituents emerges under suitable conditions. Almost all of these studies have focused on complex networks composed of classical oscillators, with the aim of predicting their synchronization behavior ~\cite{Itabashi2021,Bassi2022,Fan2022,Guth2019,Chowdhury2021,Weng2023,Zhang2022,Thiem2020,Biccari2020,Fan2021}. On the other hand, it has also been demonstrated that a quantum machine learning protocol with classical feedback can enhance synchronization between two two-level systems within two coupled cavities~\cite{Lopez2019}. Lastly, an artificial neural network has been shown to be used to infer the dissipation properties of the environment via a probe observable in an open system scenario where synchronization manifests, and the emergence of synchronization improves the performance of classification and regression tasks in this setting~\cite{Estarellas2019}.

In this work, we consider the environment-induced synchronization phenomenon, which spontaneously emerges between the expectation values of spin observables of a pair of qubits during the time evolution of the open system. We present a machine learning-based approach that employs a supervised learning algorithm to identify the synchronization behavior of the qubit pair. In particular, we utilize the $k$-nearest neighbors algorithm to identify whether the pair of qubits tends toward synchronization, antisynchronization, time-delayed synchronization, or the complete absence of these phenomena. In order to quantify the degree of synchronization between the expectation values of local qubit observables, we calculate the Pearson correlation coefficient that quantify the association between two variables that are measured in the same interval. Whereas we set the final value of the Pearson coefficient after a long time evolution as the target of our algorithm, we use the early time expectation values of the qubit observables during the dynamics as the input of the algorithm. We demonstrate the effectiveness of our approach by training and testing it with calculated quantum data obtained from three distinct open-system models, and show that it is capable of accurately predicting the synchronization behavior of the qubit pair even in the presence of random measurement errors. We establish a trade-off relation between the percentage of error in the measurement of expectation values and the number of measurements required to maintain high accuracy.

This paper is organized as follows. In Section \ref{phys_models}, we introduce the open system models that we consider. In Section \ref{data_structure}, we describe our methodology for obtaining the databases used to train and test the algorithm. In Section \ref{KNN} we provide a brief overview of how the algorithm works. In Section \ref{results}, we present the main results, demonstrating the effectiveness of our approach in predicting the behavior of the simulated quantum systems, including the addition of random errors to the measurements. Finally, in Section \ref{conclusion}, we conclude by discussing our central results and their possible applications.

\section{Physical Models} 
\label{phys_models}

In this chapter, we introduce three physical models that display distinct synchronous behaviors~\cite{Karpat2019,Karpat2020,Karpat2021}. Our main aim is to study the synchronization properties of these models from a machine learning perspective. 

\subsection{Collision Models} \label{collision_model}

Two of the models we consider in our analysis are based on collision models~\cite{Ciccarello2017,Ciccarello2021}, which are very versatile and effective tools for modeling open quantum systems. In the framework of collision models, each interaction, either between the subsystems or between the system and the environment, is described as a short unitary coupling among the involved parties, and the state of the subject system is tracked through these successive couplings. Before elaborating on the distinctive properties of the two collision models we will consider, we would like to outline the common parts. From this point on, we set $\hbar=1$ in both models, and we have two system qubits labeled as $s_1$ and $s_2$ with their free Hamiltonians
\begin{equation} \label{selfH}
H_{s_{1}}=- \frac{\omega_1}{2} \sigma^{z}_{s_1}, \quad H_{s_{2}}= -\frac{\omega_2}{2} \sigma^{z}_{s_2},
\end{equation}
where $\sigma^{z}$ is the usual Pauli operator in z-direction and $\omega_1$ and $\omega_2$ represent the self-energies of the system qubits $s_1$ and $s_2$, respectively, with their self-evolution operators, $U_{s_{1}}=\exp(-i H_{s_1} \delta t_s)$ and $U_{s_{2}}=\exp(-i H_{s_2} \delta t_s)$. In addition, we suppose that the environment in both collision models is made up of identical qubits that are all initiated in their ground state, i.e., the initial states of environment qubits $e_n, e_{n+1},...$ are $|e_n\rangle=|e_{n+1}\rangle=...=|0\rangle$, and there exist no initial correlations between the system qubits and the environment qubits. 

\begin{figure}
\centering
\includegraphics[width=0.48\textwidth]{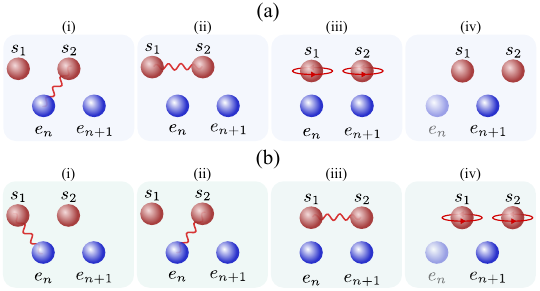}
\caption{(a) First collision model simulating local dissipation for the system qubit $s_2$: (i) The system qubit $s_2$ interacts with the environment qubit $e_n$. (ii) The system qubits $s_1$ and $s_2$ interact directly with each other. (iii) The system qubits $s_1$ and $s_2$ evolve freely. (iv) The environment qubit that has interacted with the system qubit $s_2$ in step (i) is traced out, and the same cycle is iterated with the next environment qubit $e_{n+1}$. (b) Second collision model that simulates global dissipation for system qubits $s_1$ and $s_2$: (i) The system qubit $s_1$ interacts with environment qubit $e_n$. (ii) The system qubit $s_2$ interacts with the same environment qubit $e_n$. (iii) The system qubits $s_1$ and $s_2$ directly interact. (iv) The system qubits $s_1$ and $s_2$ evolve freely, and the environment qubit that interacted with the system qubits in step (i-ii) is traced out to continue the cycle with the next environment qubit $e_{n+1}$.}
\label{fig1}
\end{figure}

Let us now continue with the specifics of the first collision model, which is schematically described in Fig.~\ref{fig1}(a). Here, the system qubit $s_2$ is coupled to the environment, while $s_1$ is isolated from it and is only allowed to interact directly with $s_2$. This model can be summarized in a few steps as follows. First, the system qubit $s_2$ interacts with the environment qubit $e_n$ through the Hamiltonian,
\begin{equation}
H_{s_2e_n}= \frac{J}{2}(\sigma^{x}_{s_2} \sigma^{x}_{e_n} + \sigma^{y}_{s_2} \sigma^{y}_{e_n}),
\end{equation}
where $\sigma^{x}$ and $\sigma^{y}$ are the standard Pauli operators in the x- and y-directions, with the corresponding unitary evolution operator given as $U_{se}=\exp(-i H_{se} \delta t_{se})$. Here, the parameter $J\delta t_{se}$ quantifies the strength of the coupling between the open system qubit $s_2$ and the environment qubits. Afterwards, the system qubits $s_1$ and $s_2$ interact with each other directly described by the Hamiltonian,
\begin{equation} \label{s1s2first}
H_{s_1s_2}= \frac{\lambda}{2}(\sigma^{x}_{s_1} \sigma^{x}_{s_2} + \sigma^{y}_{s_1} \sigma^{y}_{s_2}),
\end{equation}
where the operator $U_{ss}=\exp(-i H_{s_1s_2} \delta _{t_{ss}})$ describes their dynamics with $\lambda\delta t_{ss}$ being the coupling strength between $s_1$ and $s_2$. In the third step, the system qubits $s_1$ and $s_2$ evolve freely with their self-Hamiltonians. In the last step, a single iteration of the collision model is finalized by tracing out the environment qubit $e_n$ and moving on to repeat the procedure described above with the environment qubit $e_{n+1}$. This cycle is repeated $N$ times to obtain the evolution of the open system.

In the second collision model, different from the first model we discussed, both system qubits $s_1$ and $s_2$ are open to interaction with the environment. We describe a single cycle in the model below, which is also pictorially summarized in Fig.~\ref{fig1}(b). In the initial step, the system qubit $s_1$ interacts with the environment spin $e_n$ via the interaction Hamiltonian,
\begin{equation}
H_{s_1e_n}= \frac{J}{2}(\sigma^{x}_{s_1} \sigma^{x}_{e_n} + \sigma^{y}_{s_1} \sigma^{y}_{e_n}).
\end{equation}
Next, the system qubit $s_2$ interacts with the same environment spin $e_n$ through the same Hamiltonian, i.e.,
\begin{equation}
H_{s_2e_n}= \frac{J}{2}(\sigma^{x}_{s_2} \sigma^{x}_{e_n} + \sigma^{y}_{s_2} \sigma^{y}_{e_n}).
\end{equation}
The associated unitary evolution operator is then given by $U_{se}=\exp(-i H_{se} \delta t_{se})$ for the interaction of both system qubits with the environment, and $J\delta t_{se}$ quantifies the coupling strength. After the system-environment interactions, the system qubits $s_1$ and $s_2$ directly interact with each other with the Hamiltonian,
\begin{equation}\label{s1s2second}
H_{s_1s_2}= \frac{\lambda}{2}(\sigma^{x}_{s_1} \sigma^{x}_{s_2}),
\end{equation}
where the corresponding unitary evolution operator becomes $U_{ss}=\exp(-i H_{s_1s_2} \delta _{t_{ss}})$, and $\lambda\delta t_{ss}$ is the strength of the qubit-qubit interaction. Lastly, the system qubits $s_1$ and $s_2$ evolve freely with their self-Hamiltonians given in Eq.~(\ref{selfH}), and a single iteration in the model is concluded by tracing out the environment qubit $e_n$ to continue the dynamics with the upcoming environment qubit $e_{n+1}$.

Finally, we briefly highlight the fundamental difference between the two collision models that we consider in this work, as depicted in Fig.~\ref{fig1}. While the first collision model describes a local dissipation scenario, where only one system qubit is open to interaction with the environment, the second model lets both system qubits couple to the environmental degrees of freedom. In other words, as the first model is actually a simulation of local dissipation affecting a single qubit in the two-qubit system, the second one serves as an example of global dissipation, where both qubits are under the effect of noise. On the other hand, it is also worth noting that despite the fact that both collision models have various interaction parameters in common, the qubit-qubit interaction Hamiltonian in these models is not exactly the same, which can be seen by comparing Eq.~(\ref{s1s2first}) and Eq.~(\ref{s1s2second}). Indeed, we consider these different Hamiltonians to display the independence of our results from a particular form of qubit-qubit interaction on the prediction of synchronization based on machine learning. 

\subsection{Master Equation Model} \label{master_model}

The model we consider in this part can be employed to describe the coupling between a pair of two-level atoms and a quantized thermal electromagnetic field environment. Here, we set $\hbar=1$ and adjust the units of other parameters in accordance with this. Then, the self-Hamiltonians of the system atoms are given as $H_s=\sum_{i=1}^2\omega_i\sigma_i^z$, where $\omega_i$ represents the transition frequency between the energy levels of the $ i^{th}$ atom, and $\sigma^z$ denotes the usual Pauli operator in the z-direction. We also suppose that the system atoms have polarized dipole moments, $\mathbf{d_{eg}}$, and that they are coupled to each other via the exchange interaction Hamiltonian $H_d=\sum_ { i\neq j}^2f_{ij}\sigma^+_i\sigma^-_j$, where $f_{ ij }$ quantifies the intensity of the atom-atom interaction, and $\sigma^{\pm}$ are the up and down operators of the two-level atoms. Considering the coupling of the atoms to thermal photons and focusing on the reduced dynamics of two-level atomic systems, we obtain the well-known quantum optical master equation given by~\cite{BreuerPetruccione, Damanet2016}
\begin{equation}\label{me}
\dot{\rho} = -i\left[(H_s+H_d), \rho\right]+\mathcal{D}_-(\rho)+\mathcal{D}_+(\rho)=\mathcal{L}(\rho),
\end{equation}
where $\mathcal{D}_-(\rho)$ and $\mathcal{D}_+(\rho)$ are defined as
\begin{align}
\mathcal{D}_-(\rho)=& \sum\limits_{i,j=1}^2\gamma_{ij}(\bar{n}+1)(\sigma_j^-\rho\sigma_i^+-\frac{1}{2}\{\sigma_i^+\sigma_j^-, \rho\}), \label{emission} \\ 
\mathcal{D}_+(\rho)= & \sum\limits_{i,j=1}^2\gamma_{ij}\bar{n}(\sigma_j^+\rho\sigma_i^--\frac{1}{2}\{\sigma_i^-\sigma_j^+, \rho\}), \label{absorbtion} 
\end{align}
where $\rho$ represents the density operator for the pair of atoms. In Eq.~(\ref{me}), as the first term represents the unitary self-evolution and the exchange interaction between the atoms, the second term detailed in Eq.~(\ref{emission}) describes the spontaneous and thermally induced emission processes. Lastly, the third term given in Eq.~(\ref{absorbtion}) is responsible for the thermally induced absorption process. The rate at which these processes occur is determined by the mean number of photons $\bar{n}$ at the transition frequency. The decay rates in the quantum optical master equation are expressed as $\gamma_{ij}=\sqrt{\gamma_i\gamma_j}a(k_0r_{ij})$ where $\gamma_{i(j)}=\omega_{i(j)}^3 g$, $k_0=(\omega_1 + \omega_2) /2c$ and $g=d_{eg}^2/3\pi\epsilon_0c^3$. In addition, condition $a(k_0r_{ij})\leq 1$ needs to be satisfied to guarantee the positivity of the dynamics. Here, $d_{eg}$, c, $\epsilon_0$, and $r_{ij}$ respectively stand for the identical dipole moment of the atoms, the speed of light in vacuum, the permittivity of free space, and the relative position of the $i^{th}$ and $j^{th}$ atoms. While $a_{ij}$ controls the degree of collective behavior in the dynamics of the atoms, $f_{ij}$ quantifies the strength of the exchange interaction between them. Explicit forms of these model parameters can be found in Refs.~\cite{Damanet2016, Bhattacharya2018,Lehmberg1970}. In our analysis, we focus on the zero temperature case $\bar{n}=0$, and assume that the two-atom system undergoes fully collective dynamics, i.e., $a_{ij}=1$. 

\section{Data structure} 
\label{data_structure}

Data are a very crucial part of all machine learning techniques, since learning algorithms are trained on datasets. To train these learning models, we need a data structure that can effectively represent the information we obtain from physical simulations. Having introduced the three different open-system models that we intend to consider in our study, we will now elaborate on the formation of the databases to be used by the machine learning algorithm here to predict the synchronization behavior of the qubits. In general, data can be divided into two groups: training data and test data. While the training data is used to train the algorithm, the test data is used to evaluate its performance. As the algorithm is trained, 
it naturally has access to this part of the data and can use it to learn how to map inputs to outputs. The test data, on the other hand, are not utilized to train the algorithm. Instead, it is used to evaluate the algorithm's performance by comparing its predictions with the actual output. Within these two data groups, we can further distinguish two subgroups: features and targets. The features are the independent input variables that the algorithm uses to estimate the target, which is nothing more than the output that the model aims to predict.

In our analysis, the variables under study are the local expectation values of the system qubit observables, as we intend to explore the harmony between their dynamics to witness the emergence of spontaneous synchronization. Due to the fact that synchronization is a property of the open system dynamics rather than a particular choice of observable, it is in principle possible to choose an arbitrary qubit observable to study the synchronization properties of the system. Here, we focus on the expectation values of system qubit observables along the x direction, that is, $\langle \sigma^x_{s_1} \rangle$ and $\langle \sigma^x_{s_2} \rangle$, for concreteness. To determine the eventual synchronization behavior of the qubit pair, it is in general necessary to trace the long time dynamics of the local expectation values of qubits and thus the open system dynamics itself. However, we will only consider 5 to 100 early time expectation values for each qubit as features (inputs) of the algorithm to demonstrate the ability of our machine learning approach to successfully anticipate the long-time synchronization behavior of the open system. Moreover, as customary in the literature, we quantify the degree of synchronization between the qubits using the Pearson coefficient. Particularly, we set the Pearson coefficient as the target (output) of the learning algorithm, and calculate it in simulations for the 100 late time expectation values of the qubit observables.

Let us also introduce the definition of the Pearson coefficient and briefly elaborate on its significance in quantifying the synchronous behavior between the qubit observables. Pearson coefficient is a statistical measure that can be employed to quantify the degree of linear association between two discrete variables. It is defined as
\begin{equation}
C_{12}=\frac{\sum_{i=1}^n (x_i - \bar{x})(y_i - \bar{y})}{\sqrt{\sum_{i=1}^n ( x_i - \bar{x})^2 \sum_{i=1}^n (y_i - \bar{y})^2}}
\label{eq:pearsonCoeff}
\end{equation}
where, $\bar{x}$ and $\bar{y}$ are the average values of $x$ and $y$ respectively, and $n$ is the number of values that the variables can take. The Pearson coefficient $C_{12}$ can assume values in the interval $[-1,1]$. As the case $C_{12}=0$ indicates that the variables have no correlation, $C_{12}=1$ ($C_{12}=-1$) signifies that they are perfectly positively (negatively) correlated. In particular, while the perfect positive (negative) correlation tells us that as one variable changes so does the other variable in the same (opposite) manner. In our work, the studied variables are the expectation values of the local qubit observables $\langle \sigma^x_{s_1} \rangle$ and $\langle \sigma^x_{s_2} \rangle$. In the context of our study, we are interested in the asymptotic synchronization behavior of the qubit pair in the long time limit. Therefore, in numerical simulations, we generate the expectation values as discrete samples that cover the last 100 expectation values for each local qubit observable, also making sure that we choose a sufficiently large $n$ value for each model. In other words, whereas the asymptotic value of $C_{12}\approx0$ implies uncorrelated oscillations for the local expectation values in the long time limit, $C_{12}\approx1$ $(C_{12}\approx-1)$ means that full synchronization (anti-synchronization) is established between them. In addition, it is also possible for the Pearson coefficient $C_{12}$ to reach asymptotic values other than 1 or -1. This suggests that the mutual oscillations of the expectation values are still phase-locked in the long time limit, however, the phase difference between them is neither 0 as in full synchronization nor $\pi$ as in full anti-synchronization. Such cases are known as time-delayed synchronization~\cite{Galve2017}.

Before concluding this section, we want to discuss how we take into account potential errors in the expectation values of qubits in our simulations, which can be caused by various different sources, such as imperfect measurements or decoherence, in order to emphasize the robustness of our approach in an experimental setting. To simulate such errors, we add random errors of varying percentages to the expectation values to see whether the considered machine learning algorithm would still be able to correctly predict the values for our target in the presence of error. The error is added by summing the expectation values with a random value $\epsilon$ between -1 and 1, multiplied by an error rate between 0.005 and 0.05. This allows us to determine the accuracy of the algorithm in terms of the deviation from the calculated expectation values.

\section{$k$-Nearest Neighbors Regressor} 
\label{KNN}

Machine learning is a subfield of artificial intelligence that is concerned with developing algorithms that can learn from the available data to make predictions or decisions without being explicitly programmed to do so. The majority of learning algorithms can be classified into two categories, namely, supervised and unsupervised~\cite{Goodfellow2016}. In supervised learning, the algorithm is trained with labeled data to make predictions about future instances. These data can be either numerical or categorical. As numerical data can be used to predict continuous values, such as the price of a vehicle, categorical data can be used to classify subjects into one of several categories. Examples of supervised learning applications include image classification, speech recognition, natural language processing, and fraud detection. On the other hand, unsupervised models work with unlabeled data to explore hidden patterns that may not be directly visible~\cite{hastie2009}.

We will utilize the $k$-nearest Neighbors (KNN) algorithm, which aims to calculate the proximity between data points using a metric such as the Euclidean distance, and then estimates the output of a test sample by averaging the outputs of its $k$ nearest neighbors in the training data~\cite{alpaydin2020}. The number of neighbors K is a hyperparameter that needs to be fixed before training the model. A small value of K results in a more flexible model that can recognize complex patterns in the data, but it may possibly lead to overfitting. On the contrary, a large value of K may result in a more stable and robust model, but it may fail to capture local patterns in the data. The algorithm is typically iterated to find the optimal value of K. The KNN algorithm has several advantages, including its simplicity and interpretability, as well as its ability to work with any underlying data distribution and ease of implementation using basic math \cite{Alsharif2020}. However, it may turn out to be computationally expensive for large datasets, and careful tuning of the hyperparameters may be necessary to achieve good accuracy. The implementation of the KNN algorithm in this work has been carried out using the scikit-learn library in Python~\cite{scikit-learn}. All algorithm parameters are kept at their default settings.

\section{Results}  \label{results}

In this chapter, we present our main results related to the implementation of the KNN algorithm to predict the long-time synchronization behavior for a pair of qubits in three distinct open system models introduced previously.

\subsection{Collision Models}

\begin{figure}
\centering
\includegraphics[width=0.483\textwidth]{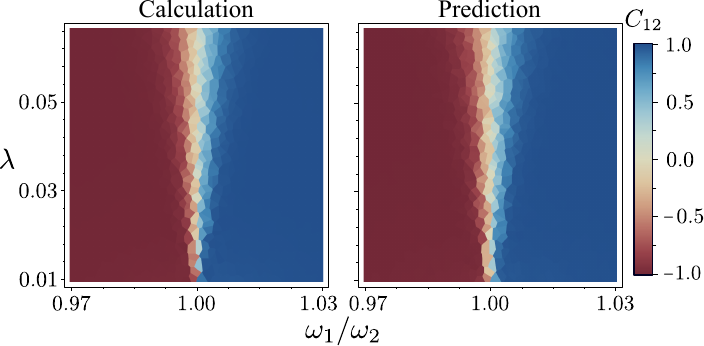}
\caption{The first collision model: (a) The Pearson coefficient $C_{12}$, calculated for the late time 100 expectation values of the system qubits $\langle \sigma^x_{s_1} \rangle$ and $\langle \sigma^x_{s_2} \rangle$, as a function of the direct interaction strength $\lambda$ between them and the detuning between their self-energies $\omega_1/\omega_2$, for the system-environment coupling strength $J=0.1$. (b) Predicted $C_{12}$ values given by the KNN regressor using the early time 100 expectation values of the qubit pair as a function of $\lambda$ and $\omega_1/\omega_2$ for $J=0.1$.}
\label{fig2}
\end{figure}

We will commence this section by considering the first collision model, outlined in Fig.~\ref{fig1}(a), where only one of the system qubits is directly coupled to the environment. We set the initial states of the system qubits $s_1$ and $s_2$ as $|\psi_{s_1}\rangle=|\psi_{s_2}\rangle=(|0\rangle+|1\rangle)/\sqrt{2}$. At this point, let us emphasize that the emergence of synchronization does not depend on the choice of initial states since it is a property of the dynamical process. Furthermore, we fix the model parameters as $\delta t_s=\delta t_{ss}=1.0$ and $\delta t_{se}=1.0$. To create our data set, we consider the values of the detuning $\omega_1/\omega_2$ varied from 0.97 to 1.03 in steps of 0.0015, the direct coupling strength $\lambda$ varied from 0.01 to 0.06 in steps of 0.002, and the system-environment interaction strength $J$ varied from 0.05 to 0.15 in steps of 0.01. This results in a total of 10,250 different configurations for the model parameters. For each of these configurations, we calculate the open system dynamics of the model and evaluate the expectation values of the observables for the system qubits, $\langle \sigma^x_{s_1} \rangle=\text{Tr}[\rho_{s_1} \sigma^x]$ and $\langle \sigma^x_{s_2}\rangle=\text{Tr}[\rho_{s_2} \sigma^x]$, after each collision, for a total of 4000 collisions, where $\rho_{s_1}$ and $\rho_{s_2}$ denote the reduced density operators for the system qubits $s_1$ and $s_2$, respectively. Confirming that the system has reached its long-time dynamical behavior with this data, we calculate the Pearson correlation coefficient $C_{12}$ by taking into account the expectation values of system qubits calculated for the last 100 collisions. In Fig.~\ref{fig2}(a), we show the Pearson coefficient $C_{12}$ plotted as a function of the detuning $\omega_1/\omega_2$ and the direct coupling strength $\lambda$ for a fixed system-environment interaction strength, that is, $J=0.1$, for presentation purposes. It is straightforward to observe that the synchronization and antisynchronization regions are separated by the resonance line on which synchronization is absent~\cite{Karpat2021}. 

To check the capability of our machine learning-based approach to predict the synchronization properties of this open system model, we calculated the corresponding expectation values $\langle \sigma^x_{s_1} \rangle$ and $\langle \sigma^x_{s_2} \rangle$  for the first 100 collisions (features), and the Pearson coefficient $C_{12}$ evaluated for the last 100 collisions (target) to train the KNN algorithm. To avoid the predictions of $C_{12}$ based directly on the training data in Fig.~\ref{fig2}, we reran the entire simulation considering slight adjustments $+\varepsilon$ in the parameters $\omega_1/\omega_2$ and $\lambda$, where $-0.0001 < \varepsilon < 0.0001$. This has in fact allowed us to train the KNN algorithm using the original dataset and evaluate its prediction performance on a new and independent dataset. In Fig.~\ref{fig2}(b), we show the predicted values of the Pearson coefficient $C_{12}$ once again, fixing $J=0.1$. However, we note that this particular choice of $J$ is made just for visualization purposes and that the KNN algorithm has actually been trained with a certain range of $J$ values. We show the calculated and predicted $C_{12}$ plots side by side in Fig.~\ref{fig2}. As can be seen by comparing the calculated and predicted Pearson coefficient plots $C_{12}$, the KNN algorithm appears to perform very well in predicting different synchronization behaviors in this model, i.e., synchronization/antisynchronization regimes and lack of synchronization. To quantify the accuracy of the algorithm, we calculate the mean absolute error (MAE) for the predicted values as compared to the calculated ones using the expression
\begin{equation}
MAE = \frac{1}{n}\sum_{i=1}^{n}\left|y_i - \hat{y}_i\right|,
\end{equation}
where $n$ is the number of data points, $y_i$ is the calculated $C_{12}$ value, and $\hat{y}_i$ is the predicted $C_{12}$ value. That is, MAE is simply given by the absolute value of the difference between the calculated and predicted values averaged over all data points. For the prediction results displayed in Fig.~\ref{fig2}, we find $MAE=0.009$, which exhibits the success of the KNN algorithm in inferring the long-time synchronization behavior of the model even though it only has access to early time expectation values for the system qubits calculated for the first 100 collisions.

\begin{figure}
\centering
\includegraphics[width=0.37\textwidth]{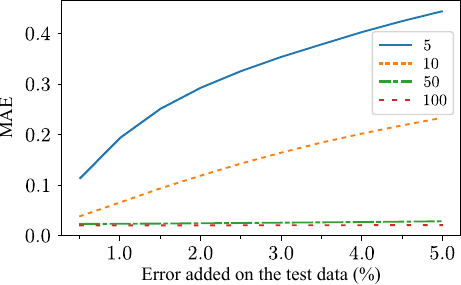}
\caption{Mean absolute error (MAE) for the Pearson coefficient $C_{12}$ in the first collision model in terms of the percentage error that is added to the calculated expectation values of the qubits $\langle \sigma^x_{s_1} \rangle$ and $\langle \sigma^x_{s_2} \rangle$. Four different lines in the legend correspond to the cases of feeding the KNN regressor with 5, 10, 50, and 100 pairs of qubit expectation values for prediction.}
\label{fig3}
\end{figure}

Moreover, to examine a more realistic scenario, we take into account potential errors in the measurement of the local expectations values of the qubits, which can naturally occur in an experimental setting. As mentioned previously, we simulate these errors by summing the expectation values with a random number $\epsilon$ taken from the interval $\epsilon\in[-1,1], $ which is then multiplied by an error rate between 0.005 and 0.05. In other words, we consider a percentage error rate between $\%0.5$ and $\%5$ in the expectation value measurements of both qubits. We also perform the same analysis displayed in Fig.~\ref{fig2}, where the first 100 expectation values are used as the input of the KNN regressor, but this time using the first 5, 10 and 50 expectation values of the qubit pair as the input of the algorithm. We summarize the outcomes of our simulations in Fig.~\ref{fig3}, where the dotted red, dot-dashed green, dashed orange, and solid blue lines display how the MAE for the Pearson coefficient $C_{12}$ scales with the added percentage error to the expectation values of qubits, $\langle \sigma^x_{s_1} \rangle$ and $\langle \sigma^x_{s_2} \rangle$, when the expectation values after the first 100, 50, 10 and 5 collisions are used for prediction, respectively. Based on these results, it is quite clear that there exists a trade-off relation between the number of expectation values of the qubits used for the prediction of $C_{12}$ and the robustness of the KNN regression algorithm against errors. Particularly, the predictions of the algorithm regarding the synchronization properties of the model are quite susceptible to even a small percentage of errors when its input involves a small number of qubit expectation values, as can be seen from the case of 5 expectation values for each qubit shown by the solid blue line in Fig.~\ref{fig3}. On the other hand, we also observe that once the algorithm has access to 50 or 100 expectation values for each qubit, then it can actually make quite accurate predictions of the Pearson coefficient $C_{12}$. Indeed, in such cases, increasing the percentage error in the expectation values from $0.5\%$ up to $5\%$ only very marginally changes the MAE, which is not even visible to the eye in Fig.~\ref{fig3}. All in all, these results suggest that, in the case of the first collision model describing the local dissipation of one of the qubits in the pair, our machine learning approach based on the KNN regression algorithm is quite effective in predicting different regimes of synchronization or its absence accurately. We should lastly emphasize once again that 50 to 100 expectation value measurements for each qubit at the early stages of the dynamics turn out to be sufficient to predict the long time synchronization behavior of this model with high accuracy, despite the presence of potential measurement errors in a given experimental setup.

\begin{figure}
\centering
\includegraphics[width=0.483\textwidth]{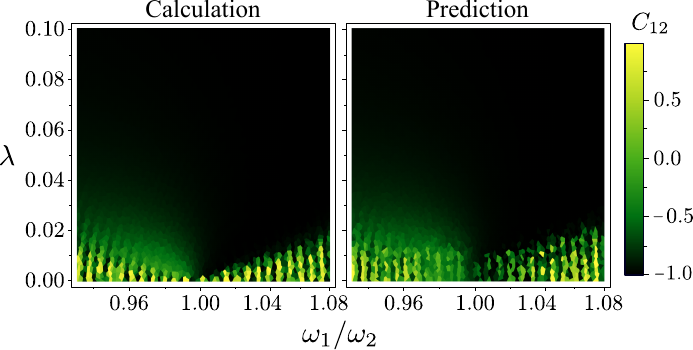}
\caption{The second collision model: (a) The Pearson coefficient $C_{12}$, calculated for the late time 100 expectation values of the system qubits $\langle \sigma^x_{s_1} \rangle$ and $\langle \sigma^x_{s_2} \rangle$, as a function of the direct interaction strength $\lambda$ between them and the detuning between their self-energies $\omega_1/\omega_2$. (b) Predicted $C_{12}$ values given by the KNN regressor using the early time 100 expectation values of the qubits as a function of $\lambda$ and $\omega_1/\omega_2$.}
\label{fig4}
\end{figure}

Having demonstrated the performance of our approach in the first collision model, we now turn our attention to the second collision model depicted in Fig.~\ref{fig1}(b), which describes the global dissipation of the qubit pair in the sense that both qubits interact with the same environmental particle at each iteration. Similarly to the first collision model, we set the initial states of the system qubits $s_1$ and $s_2$ as $|\psi_{s_1}\rangle=|\psi_{s_2}\rangle=(|0\rangle+|1\rangle)/\sqrt{2}$, and choose the model parameters as $\delta t_s=\delta t_{ss}=0.2$, $\delta t_{se}=0.05$ and $J=1.0$. After fixing the model and system parameters, we create our data set considering the values of detuning $\omega_1/\omega_2$ from 0.93 to 1.08 in steps of 0.002 and the direct coupling strength $\lambda$ from 0.0 to 0.01 in steps of 0.002. This gives us a total of 3876 different configurations for the model parameters. We obtain the open-system evolution of the model and the expectation values $\langle \sigma^x_{s_1} \rangle$ and $\langle \sigma^x_{s_2} \rangle$ for each of these configurations for 5000 collisions, after which point the expectation values become negligibly small. Then, we calculate the Pearson coefficient $C_{12}$ for the expectation values of the system qubits considering the last 100 collisions to see the long-time synchronization behavior. In Fig.~\ref{fig4}(a), the Pearson coefficient $C_{12}$ is plotted as a function of the direct coupling strength $\lambda$ and the detuning $\omega_1/\omega_2$ between the self-energies of the qubits. Unlike the first collision model, where both  synchronization and  anti-synchronization can occur depending on the detuning between the self-energies of the qubits, here it is only possible to observe anti-synchronization, which emerges if the direct coupling between the qubit pair is sufficiently strong to compensate for the detuning.

\begin{figure}
\centering
\includegraphics[width=0.38\textwidth]{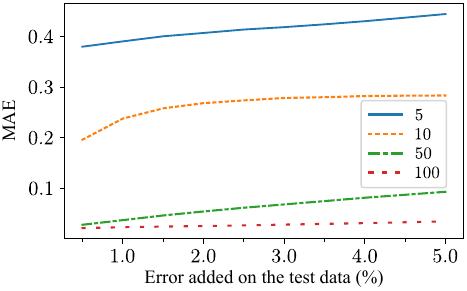}
\caption{Mean absolute error (MAE) for the Pearson coefficient $C_{12}$ in the second collision model as a function of the percentage error added to the calculated expectation values $\langle \sigma^x_{s_1} \rangle$ and $\langle \sigma^x_{s_2} \rangle$. Four different lines in the legend correspond to the cases of feeding the KNN regressor with 5, 10, 50, and 100 pairs of qubit expectation values for prediction.}
\label{fig5}
\end{figure}

Using the same approach as we followed in the examination of the first collision model, we train the KNN regressor with the corresponding expectation values $\langle \sigma^x_{s_1} \rangle$ and $\langle \sigma^x_{s_2} \rangle$ calculated for the first 100 collisions, and the Pearson coefficient $C_{12}$ evaluated for the last 100 collisions. In Fig.~\ref{fig4}(b), we show the Pearson coefficient $C_{12}$ predictions performed by the KNN algorithm. We should also emphasize that we prevent the regressor from having access to the training data by re-adjusting it with small deviations $\varepsilon$ from the model parameters $\lambda$ and $\omega_1/\omega_2$, as has been done for the case of the first collision model. Directly comparing the plots for the calculated and predicted $C_{12}$ values side by side in Fig.~\ref{fig4}, it can be observed that the KNN regression algorithm achieves a high accuracy in predicting the long time synchronization properties of the considered model. Indeed, MAE in this case turns out to be 0.040. Lastly, we consider possible measurement imperfections for the expectation values of the qubits. Taking into account random percentage errors in the qubit expectation values $\langle \sigma^x_{s_1} \rangle$ and $\langle \sigma^x_{s_2} \rangle$ between $\%0.5$ and $\%5$, we repeat the same investigation to understand how resilient our machine learning approach, based on the KNN regressor, is against potential errors in algorithm input. We present the results of our calculations in Fig.~\ref{fig5}, where the first 5, 10 and 50 expectation values of the qubit pair are considered as an input in addition to the case of the first 100 expectation values. In Fig.~\ref{fig5}, the dotted red, dot-dashed green, dashed orange, and solid blue lines demonstrate how the MAE for $C_{12}$ changes with the percentage error in the expectation values of the qubit pair when the first 100, 50, 10 and 5 expectation values are used for prediction, respectively. We observe that the KNN regression algorithm can provide reliable predictions regarding synchronization also in the second collision model, even in the presence of measurement errors in a realistic setting provided that the number of expectation values for each qubit, which are used as input variables, is sufficient.

To summarize this subsection, we see that our machine learning approach to the prediction of synchronization in open system dynamics proves to be quite successful in two distinct models. That is, in both local and global dissipation regimes described within the collision model framework, the KNN regressor can accurately foretell the emergence of (anti-)synchronization or its absence in the long time dynamics, only using the early time expectation values of the considered qubit observables.

\begin{figure}[t]
\centering
\includegraphics[width=0.483\textwidth]{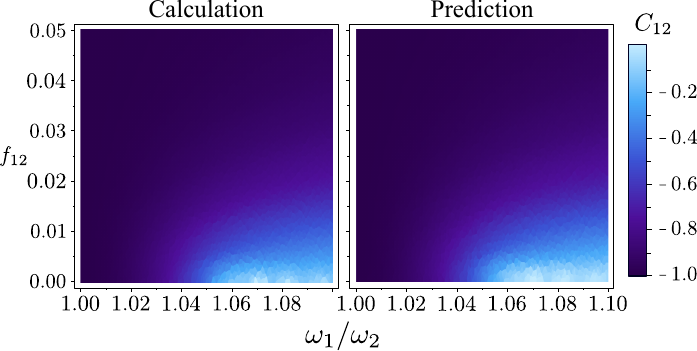}
\caption{The master equation model: (a) The Pearson coefficient $C_{12}$, calculated for the late time 100 expectation values of the qubits $\langle \sigma^x_{s_1} \rangle$ and $\langle \sigma^x_{s_2} \rangle$, as a function of the exchange interaction strength $f_{12}$ between them and the detuning between their self-energies $\omega_1/\omega_2$. (b) Predicted $C_{12}$ values produced by the KNN regressor using the early time 100 expectation values of the qubit pair in terms of $f_{12}$ and $\omega_1/\omega_2$.}
\label{fig6}
\end{figure}

\subsection{Master Equation Model}

Up to this point, we have considered two open-system models which represent local and global dissipation scenarios, making use of the collisional model framework. We have seen that our machine learning-based approach to the detection of the dynamical emergence of synchronization performs quite accurately in both models despite potential experimental errors in the qubit expectation value measurements. As a last test, we consider an open-system model described by a master equation, where two two-level atomic particles (or qubits) interact with an electromagnetic field fully collectively. In this setting, depending on the detuning between the self-energies of the qubits and the direct coupling between them, it is possible to observe phase-locked oscillations of expectation values, where the phase difference is not necessarily $\pi$ or 0 degrees, i.e., time-delayed synchronization~\cite{Karpat2020}.

Let us first recall that we will consider the collective dissipation of the qubit pair in the zero temperature limit, hence we fix $\bar{n}=0$ and $a_{ij}=1$. Also, we set the initial states of the system qubit pair as $|\psi_{s_1}\rangle =(|0\rangle+|1\rangle)/\sqrt{2}$ and $|\psi_{s_2}\rangle =(|0\rangle+e^{-i \pi /3}|1\rangle)/\sqrt{2}$. To create our data set, we let the detuning between the self-energies of the qubits $\omega_1/\omega_2$ acquire values between 1.0 and 1.1 in steps of 0.001, and the strength of the exchange interaction between the qubits $f_{12}$ between 0 and 0.05 in steps of 0.001. As a result, we obtain 5151 different model parameter configurations, for each one of which we simulate the open system dynamics and then calculate the expectation values $\langle \sigma^x_{s_1} \rangle$ and $\langle \sigma^x_{s_2} \rangle$ until $t=500$. We note that at this point in the dynamics, the system already displays its long time behavior, and here we simulate the time evolution in discrete time steps, i.e., taking $t=0, 1,...,500$. Next, we evaluate the Pearson coefficient $C_{12}$ for the expectation values of the qubit pair observables, taking into account the last 100 time steps in the dynamics, which shows us the long time synchronization behavior of the open system model. We demonstrate the outcomes of our calculations in Fig.~\ref{fig6}(a). It is straightforward to notice that when the exchange interaction between the qubits is sufficiently strong to compensate for the detuning between their self-energies, full anti-synchronization emerges between the expectation values of the qubit observables. In addition, for this model, it is known that even if the expectation values do not get fully anti-synchronized, there still occurs time-delayed synchronization~\cite{Karpat2020}. In other words, even if the Pearson coefficient $C_{12}\neq -1$ in a long time, it still settles to a constant value between 0 and -1 as shown in Fig.~\ref{fig6}(a), which signals the onset of oscillations having a fixed phase difference between the expectation values.

\begin{figure}
\centering
\includegraphics[width=0.38\textwidth]{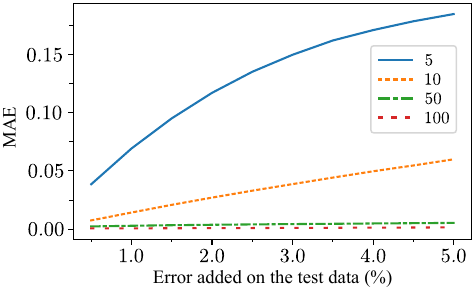}
\caption{Mean absolute error (MAE) for the Pearson coefficient $C_{12}$ in the master equation model in terms of the added percentage error to the calculated expectation values of the qubits $\langle \sigma^x_{s_1} \rangle$ and $\langle \sigma^x_{s_2} \rangle$. Four different lines in the legend correspond to the cases of feeding the KNN regressor with 5, 10, 50 and 100 pairs of qubit expectation values for prediction.}
\label{fig7}
\end{figure}

Finally, we will check the ability of the KNN regressor to predict the synchronization behavior of the qubit pair for the considered master equation model. As has been done for the collision models, we feed the KNN regressor with the dataset for this configuration, which involve the qubit expectation values $\langle \sigma^x_{s_1} \rangle$ and $\langle \sigma^x_{s_2} \rangle$ calculated at the first 100 time points, and the Pearson coefficient $C_{12}$ evaluated considering the last 100 time points. The predicted $C_{12}$ values produced by the KNN regression algorithm are shown in Fig.~\ref{fig6}(b) side by side with the calculated $C_{12}$ values. It is obvious from the comparison of the plots that our approach based on the KNN regressor performs excellently in predicting the synchronization properties of the model. In fact, the MAE in this case is calculated to be only 0.002. To complete this section, we now consider the potential measurement inaccuracies in the expectation values to assess the robustness of our machine learning-based approach against errors. Similarly to the previously studied models, we randomly add percentage errors to the expectation values of the qubit pair between $0.5\%$ and $5\%$, and perform the same analysis by considering 5, 10, 50, and 100 early time expectation values as the input of the regressor. Fig.~\ref{fig7} displays the results of this investigation, where the dotted red, dot-dashed green, dashed orange, and solid blue lines show how the MAE for the Pearson coefficient $C_{12}$ changes as a function of the percentage error added to the expectation values $\langle \sigma^x_{s_1} \rangle$ and $\langle \sigma^x_{s_2} \rangle$, when the expectation values at the first 100, 50, 10 and 5 time points are given to the regressor for prediction, respectively. It is evident from Fig.~\ref{fig7} that as long as 10 or more early time expectation values for the qubit pair are fed to the algorithm as input, the KNN regressor does an excellent job predicting the late time synchronization behavior of the model despite the presence of errors in the expectation values.

\subsection{Possible Physical Error Mechanisms}

In this section, we will investigate possible error mechanisms that could affect the state of the quantum system in question, prior to the prediction stage. We will assume that our algorithm is trained with the data obtained from the ideal case and try to assess the impact of such potential errors as quantified via the MAE.

\begin{figure}[t]
\centering
\includegraphics[width=0.8\columnwidth]{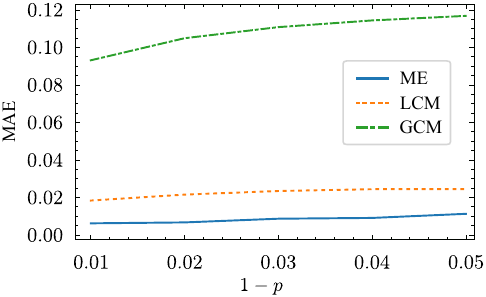}\\
\caption{Mean absolute error (MAE) for the Pearson coefficient $C_{12}$ in terms of the error probability in the initial state that is fed into the KNN regressor for the master equation model (ME), local collision model (LCM), and global collision model (GCM) using 100 early time expectation values.}
\label{fig8}
\end{figure}

First, we intend to examine the effects of initialization errors, which might occur in experiments, on the initial state of the system. To account for realistic conditions, we now consider a modified initial state for all three models that is described by the density matrix, 
\begin{equation}
\rho = p|++\rangle\langle++| + (1-p)\left(|+-\rangle\langle+-| + |-+\rangle\langle-+|\right).
\end{equation}
This modification accommodates the potential of qubit states being incorrectly initialized through two specific mechanisms: a phase error where an intended $|+\rangle$ state initializes as $|-\rangle$, or a bit-flip error since the state $|+\rangle$ can be generated from the $|0\rangle$ state via a Hadamard gate. In the latter case,  $1-p$ reflects the probability that the initialization intended for the state $|0\rangle$ actually ended up to give the state $|1\rangle$ , leading to the unintended creation of a $|-\rangle$ state by the Hadamard transformation. Such errors are relatively common, particularly in superconducting qubit platorms, where there is always a non-zero probability of measuring the state $|1\rangle$ ($|0\rangle$) when the intention was to create the state $|0\rangle$ ($|1\rangle$) \cite{ibmquantum}. In Fig.~\ref{fig8}, we present the MAE as a function of $1-p$, i.e., the probability of having an initialization error, which shows us how sensitive our method is to a possible initial state error in the prediction stage for the three different models.

In all of the cases investigated in this subsection, we calculate the expectation values using 100 
time points. We also note that here we refer to the first and second collision models as local collision model (LCM) and global collision model (GCM) for presentation purposes. We observe in Fig.~\ref{fig8} that the master equation (ME) model demonstrates the least sensitivity to initialization errors, maintaining a consistently low MAE, with an increase from 0.002 (in the absence of initialization errors) to values still below 0.01. In contrast, LCM and GCM exhibit higher MAEs, with the LCM's MAE increasing from 0.009 to approximately 0.02 and the GCM's MAE rising from 0.04 to the order of 0.1. Despite these increases, the overall changes in MAE remain modest, which shows the robustness of our method against initialization errors. Notably, the GCM experiences a relatively more significant effect, with MAE now in the order of 10\% when considering initialization errors. Such errors still remain relatively low, highlighting the efficiency of our approach.

Furthermore, we have explored how the temperature of the environment influences the ability of the algorithm to predict the emergence of (anti-)synchronization across all three considered models. To this aim, we have evaluated the prediction performance of the KNN regressor in the presence of non-zero environmental temperatures, when the algorithm is still trained using the zero temperature data. In order to take into account the temperature effects, we assume that the density matrices of the environmental qubit in both local and global collision models are given by $\rho_E = (1-p)|0\rangle\langle0| + p|1\rangle\langle1|$, with $p$ ranging from $0.01$ to $0.05$. Physically, this implies that the environment is actually found in a mixed thermal state, reflecting the effects of the temperature. In particular, as $p$ changes from $0.01$ to $0.05$, a gradual increase in the environmental temperature is being simulated, where the probability of the environment being in the ground state decreases while at the same time its probability of being in the excited state increases. On the other hand, in case of the master equation model, we slightly vary the average number of photons, allowing $\bar n$ to change from $0.01$ to $0.05$ in Eq.~(\ref{emission}) and Eq.~(\ref{absorbtion}), which lets us analyze the effect of the temperature of the environment on the performance of the KNN regressor.

\begin{figure}[t]
\centering
\includegraphics[width=0.8\columnwidth]{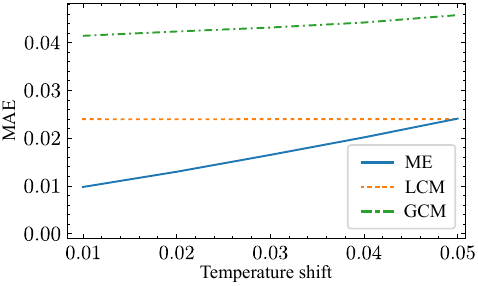}\\
\caption{Mean absolute error (MAE) for the Pearson coefficient $C_{12}$ as a function of the temperature of the environment for the master equation model (ME), local collision model (LCM), and global collision model (GCM) considering 100 early time qubit expectation values.}
\label{fig9}
\end{figure}

In Fig.~\ref{fig9}, we display the MAE for our predictions obtained utilizing the KNN regressor, trained with the ideal dataset, as a function of the environment temperature for all three models. We observe that for the GCM and ME, the MAE slightly increases as the environment temperature rises. Nonetheless, in case of the LCM, where one of the qubits is not directly in contact with the environment, the performance of our approach is basically indifferent to the changes in the environment temperature.

\section{Conclusion} \label{conclusion}

To sum up, we have employed a machine learning technique, based on the KNN regression algorithm, to predict the emergence of environment-induced spontaneous synchronization between the expectation values of observables for a pair of qubits in the dynamics of open quantum systems. We have considered three different open system models that cover local, global, and collective dissipation regimes. While the first two models are described within the collisional model framework, the third one is represented by a master equation of Lindblad form. We have demonstrated that in all of these three models, our machine learning-based approach is able to identify the long time synchronization behavior of the open system with quite high accuracy, only utilizing the early time expectation values for the qubit observables as an input. In fact, we have shown that the machine learning algorithm not only very accurately estimates the onset of synchronization or anti-synchronization between the qubit expectation values but also the phase-locked oscillations between them with arbitrary phase difference, i.e., time-delayed synchronization. The proposed approach has the potential to be useful in experiments since it can significantly reduce the number of measurements required to determine the dynamical establishment of synchronous evolution. To see the robustness of our approach against potential measurement imperfections in an experimental setting, we added random errors to the simulated qubit expectation values and benchmarked the capability of our method to recognize the synchronization behavior in the models considered. Additionally, we have tested the robustness of our model to small variations in the initial state of the system, as well as to temperature, showing that despite a slight increase in the MAE, the predictions remained fairly accurate. We have established a trade-off relation between the degree of randomly introduced error in the expectation values and the accuracy of the algorithm to predict synchronization. All in all, our results demonstrate that despite the presence of errors in the input values of the KNN regressor, the proposed approach still works with high accuracy as long as a few dozen early time qubit expectation values are introduced to the algorithm for prediction. The fact that the employed method works accurately in three different open system scenarios suggests that it might actually be applied in a wide range of physical models efficiently. Lastly, motivated by the performance of the employed machine learning approach in succesfully predicting the appearance of mutual synchronization between two qubits, one can  try to apply similar methods to more sophisticated problems involving a higher number of principal quantum spins or oscillators, where exotic dynamical states such as chimera states can emerge under suitable conditions~\cite{Bastidas2015,Senthilkumar2019}.

\begin{acknowledgments}
F. M. acknowledges support from Coordena\c{c}{\~a}o de Aperfei\c{c}oamento de Pessoal de N{\'i}vel Superior (CAPES), project number 88887.607339/2021-00. B. \c{C}. and G. K. are supported by The Scientific and Technological Research Council of Turkey (TUBITAK) under Grant No. 121F246. F.F.F acknowledge support from Funda\c{c}{\~a}o de Amparo {\`a} Pesquisa do Estado de S{\~a}o Paulo (FAPESP), project number 2023/04987-6 and from ONR, project number N62909-24-1-2012. \.{I}. Y. acknowledges financial support from Czech Grant Agency project number GAČR 23-07169S.
\end{acknowledgments}

\bibliography{bibl}

\newpage

\onecolumngrid
\appendix

\section{Early time behavior and emergence of synchronization}\label{appdxA}

\begin{figure}[h]
\centering
\includegraphics[width=0.6\columnwidth]{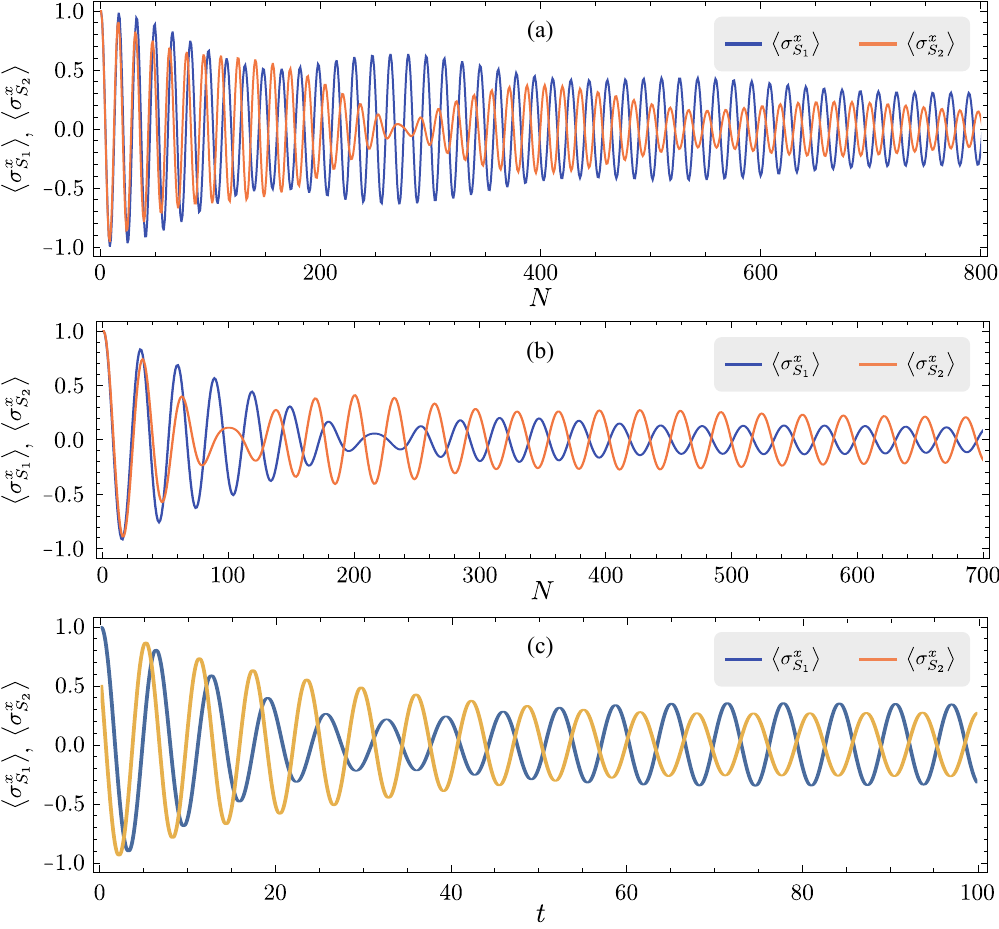}
\caption{Dynamics of the local spin expectation values of the two qubits for the local collision model (LCM) (upper), global collision model (GCM) (middle), and master equation model (ME) (lower), which display both their pre-synchronous dynamics and also the emergence of dynamical transition into the (anti-)synchronous behavior.}
\label{fig10}
\end{figure}

We present the explicit dynamics of the spin expectation values for all three models we consider in the manuscript in Fig.~\ref{fig10}. All subfigures include pre-synchronization dynamics together with the emergence of (anti-)synchronized dynamics in all three models. Once the synchronous behavior is established between the observables, it persists either in the steady-state or until the expectation values decay to zero. The time window that we use to train the machine learning algorithm consists of the data after the synchronization is established. Then, using this algorithm, we try to predict synchronization by looking at the very early time data (up to first 100 points) where the synchronization is not yet established, as outlined in Sec.~\ref{data_structure}. Therefore, we believe that our approach to the problem is a promising way to guess the late-time behavior of a property of a quantum system by looking at the early-time dynamics.

\vspace{1cm}

\end{document}